\documentclass[prb,preprint,groupedaddress,showpacs]{revtex4}
\usepackage{graphicx} 
\begin{document} 
\bibliographystyle{prbrev}
 
\title{Optical response of two-dimensional few-electron concentric double 
quantum rings: A local-spin-density-functional theory study}

\author{F. Malet}
\affiliation{Departament ECM, Facultat de F\'{\i}sica,
and IN$^2$UB,
 Universitat de Barcelona. 
Diagonal 647, 08028 Barcelona, Spain}

\author{M. Pi}
\affiliation{Departament ECM, Facultat de F\'{\i}sica,
and IN$^2$UB,
Universitat de Barcelona. Diagonal 647,
08028 Barcelona, Spain}

\author{M. Barranco}
\affiliation{Departament ECM, Facultat de F\'{\i}sica,
and IN$^2$UB,
Universitat de Barcelona. Diagonal 647,
08028 Barcelona, Spain}

\author{E. Lipparini}
\altaffiliation{Permanent address:
Dipartimento di Fisica, Universit\`a di Trento, and INFN,
38050 Povo, Trento, Italy}
\affiliation{Departament ECM, Facultat de F\'{\i}sica,
and IN$^2$UB,
Universitat de Barcelona. Diagonal 647,
08028 Barcelona, Spain} 

\author{Ll.\ Serra}
\affiliation{Departament de F\'{\i}sica, Universitat de les Illes Balears,
and Institut Mediterrani d'Estudis Avan\c{c}ats IMEDEA (CSIC-UIB),
 E-07122 Palma de Mallorca, Spain}

\date{\today}

\begin{abstract}
We have investigated the dipole charge- and spin-density response of
few-electron two-dimensional concentric nanorings as a function of the
intensity of a perpendicularly applied magnetic field.
We show that the dipole response displays signatures associated with
the localization of electron states in the inner and outer ring favored
by the perpendicularly applied magnetic field. Electron localization
produces a more fragmented spectrum due to the appearance of
additional edge excitations in the inner and outer ring.

\end{abstract}

%
%

\maketitle

Progress in nanofabrication technology has allowed to
produce self-assembled, strain-free, nanometer-sized quantum complexes
consisting of two concentric,
well-defined GaAs/AlGaAs rings\cite{Man05,Kur05} whose theoretical study
has recently
attracted some interest.\cite{Fus04,Sza05,Pla05,Cli06} 
Most of these
works
are concerned with the properties of their ground state, although
the optical properties of one- and two-electron concentric double
quantum
rings (CDQR) at zero magnetic field have been addressed using a
single-band effective-mass envelope
model discarding Coulomb correlation effects.\cite{Kur05}

The singular geometry of CDQR has been found to introduce characteristic
features in the addition spectrum compared to that of other coupled
nanoscopic quantum structures. As a function of the
inter-ring distance, the localization of the electrons in either ring
follows from the interplay between
confining, Coulomb and centrifugal energies. Each of them
prevail in a different range of inter-ring distances, affecting in a
different way the CDQR addition spectrum.\cite{Cli06}
It is thus quite natural to investigate whether and how  
localization effects may show up in the dipole response, using a
perpendicularly applied magnetic field instead of the
inter-ring distance to control the electron localization in either
ring.

The aim of this paper is to use local-spin-density-functional
theory (LSDFT) as described in detail in Ref. \onlinecite{Ser99},
to investigate the dipole longitudinal response of CDQR. The method has
been used in the past to address the response of
single quantum rings (see e.g. Ref. \onlinecite{Emp01} and references
therein). We address here the  few-electron case, and 
consider the CDQR's as strictly two-dimensional systems. 

Within LSDFT, the ground state  of the system is
obtained by solving the Kohn-Sham (KS) equations.
The problem is simplified by the imposed circular symmetry
around the $z$ axis, which allows one to write the
single particle (sp) wave functions as
$\varphi_{nl\sigma}({\bf r},\sigma)=
u_{nl\sigma}(r) e^{-\imath l \theta} \chi_{\sigma}$,
being $-l$
the projection of the sp orbital angular momentum on the $z$ axis.
The confining potential has been taken in a form that
slightly generalizes that of Ref. \onlinecite{Sza05}:
$$V_{cf}(r) =  {\rm min} \left\{
\frac{1}{2}\, m \,\omega_1^2\, (r - R_1)^2 \; , \;
\frac{1}{2}\, m \,\omega_2^2\, (r - R_2)^2 \right\} \;\; , $$
with $R_1=20$ nm, $R_2=40$ nm, $\omega_1=30$ meV, and $\omega_2=40$ meV.
The radii have been fixed to the experimental values,\cite{Kur05} while
the frequencies are rather arbitrary. We have considered
large frequencies to mimic the strong confinement
felt by the CDQR, and have taken $\omega_2 > \omega_1$ to
somewhat compensate that, as $R_2 >> R_1$, the `surface' of the outer
ring might have been overestimated if we had taken both frequencies
equal. In three dimensions, more realistic
confining potentials, better suited to model the
experimental devices,\cite{Kur05} have been considered.\cite{Cli06}
On the CDQR system it may act a constant magnetic field $B$ in
the $z$ direction, to which a cyclotron frequency $\omega_c = e B/mc$ 
is associated. 
We have taken for the dielectric constant, electron effective mass,
and effective gyromagnetic factor, the values appropriate for GaAs,
namely, $\epsilon$ = 12.4,   $m^*$ = 0.067, and
$g^*=-0.44$, and have solved the KS equations for up to $N=6$ 
electrons, and for $B$ values up to 4-5 T, depending on $N$.
In the following, we discuss some illustrative results.

\begin{figure}[t]
\centerline{\includegraphics[width=6.5cm,clip]{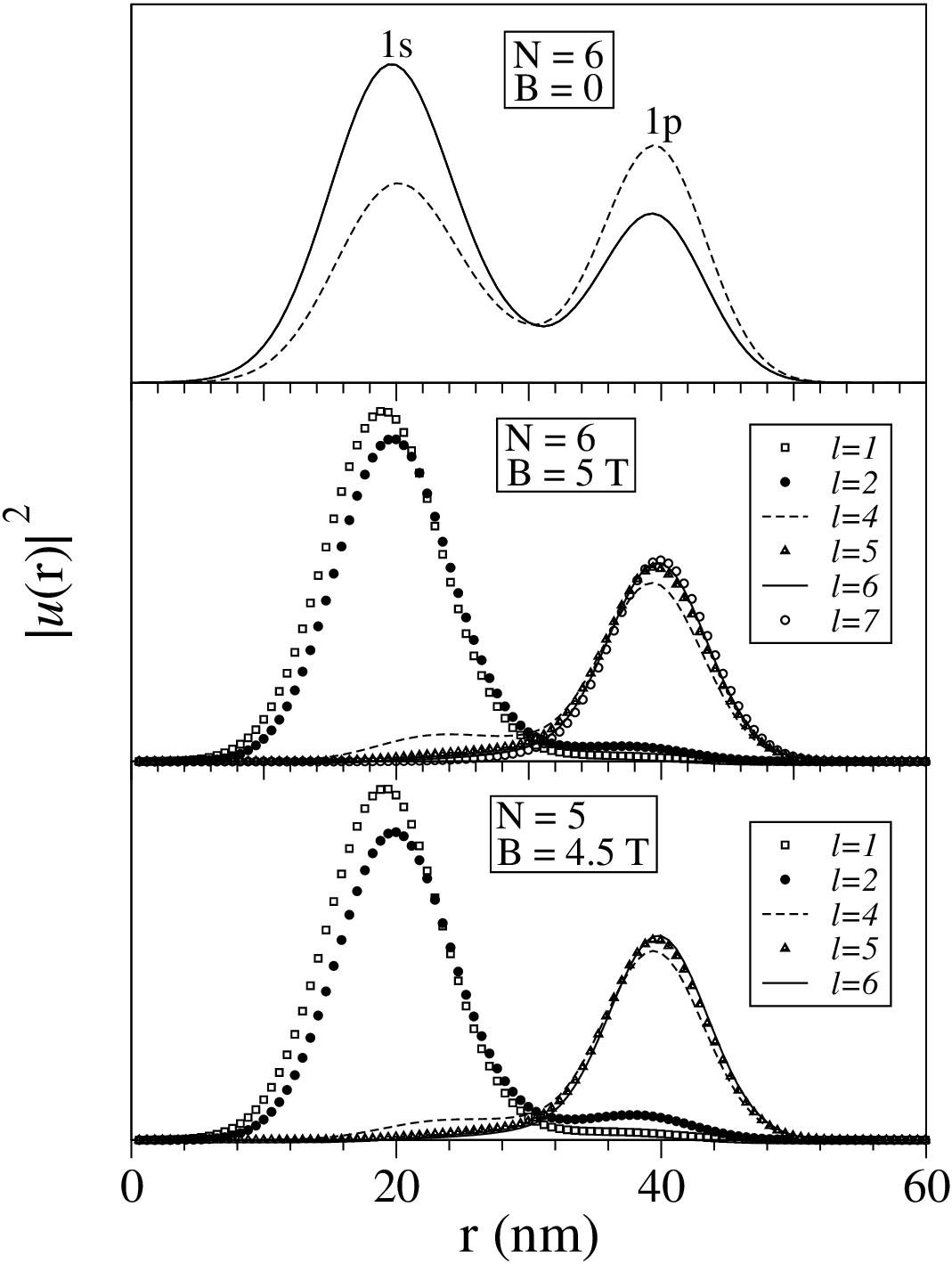} }
\caption{Top panel: Squared wave functions
(arbitrary units) as a function of $r$ (nm) of the
KS occupied orbitals corresponding to $N=6$,
$B=0$ T.
Middle panel: same as top panel for $N=6$, $B=5$ T;
all orbitals are spin polarized. Notice that the $l=3$ orbital is 
not occupied.
Bottom panel: same as middle panel for $N=5$, $B=4.5$ T.
}
\label{fig1}
\end{figure}

Fig. \ref{fig1} shows the squared wave functions $|u(r)|^2$
corresponding to the KS occupied orbitals for $N=6$ and $B=0$ T
(top panel). In this
case, the total spin of the CDQR is zero, and the spin-up and down
orbitals corresponding to the same $(n,|l|)$ values are degenerate.
As a consequence, there are only two different radial wave functions,
one for $s\,(l=0)$ states, and another for $p\,(l=\pm 1)$ states. It can
be seen that electrons are fairly delocalized within the CDQR; for
the chosen confining potential, localized configurations would only 
appear at larger inter-ring distances.\cite{Cli06}

It is known that for an $N$-electron  single quantum ring,
sp states with small $l$ values become progressively empty as $B$
increases.
This can be intuitively understood as follows. If
only nodeless radial states are occupied, the Fock-Darwin  wave
functions are proportional to $x^{|l|} e^{- x^2/4}$, where $x=r/a$,
being $a=\sqrt{\hbar/(2 m \Omega)}$ with $\Omega=\sqrt{\omega_0^2+
\omega_c^2/4}$. Of course, this is so for a harmonic confining
potential of frequency $\omega_0$, and not for the ring confining
potential, but some of that wave function structure remains even in this
case. These wave functions are peaked at 
$r_{max} \sim \sqrt{ 2 |l|}\,a$,
and consequently, as $B$ increases, the $|l|$ values corresponding to
occupied levels must increase so that $r_{max}$ sensibly lies within 
the range of $r$ values spanned by the ring morphology.
The same appears to happen for the CDQR we have studied, as 
illustrated in Fig. \ref{fig1}.
The squared wave functions $|u(r)|^2$ for $N=6$,  $B=5$ T, and
for $N=5$, $B=4.5$ T are displayed in the middle and
bottom panels of this figure, respectively, showing
that indeed, high-$l$ orbitals are mostly localized in the outer
ring, whereas low-$l$ orbitals are mostly localized in the inner ring.
A complete localization of all orbitals could have been achieved using
e.g. a larger $\omega_1$ value. 

\begin{figure}[t]
\centerline{\includegraphics[width=6.5cm,clip]{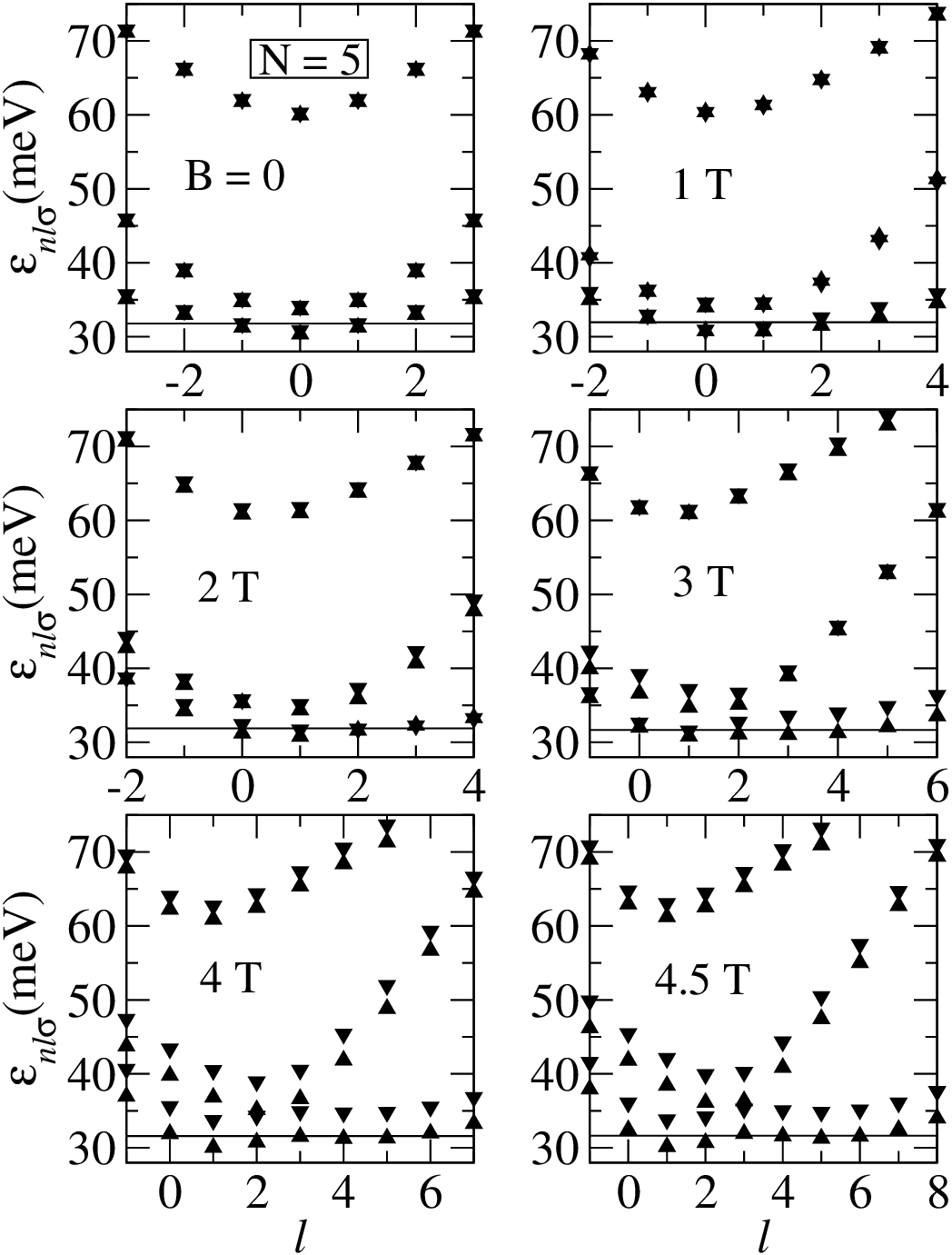} }
\caption{ Kohn-Sham sp energies (meV) of the $N=5$ CDQR ground
state
as a function of $l$ for several $B$ values. Upward(downward)
triangles correspond to
spin-up(down) orbitals. The thin horizontal line indicates the
Fermi level.}
\label{fig2}
\end{figure}

\begin{figure}[t]
\centerline{\includegraphics[width=7cm,clip]{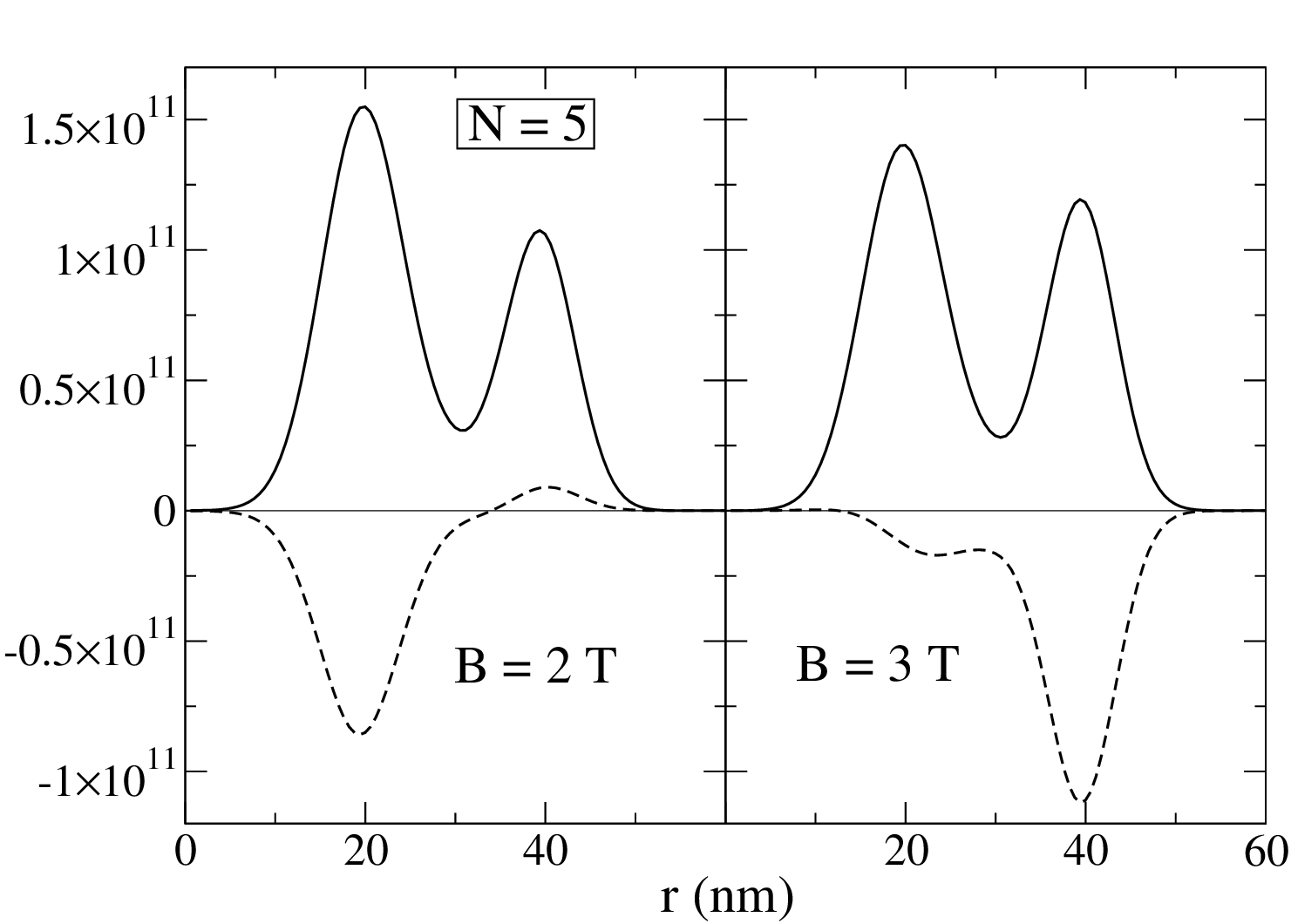} }
\caption{ Electron density $n(r)$ (solid line) and spin magnetization 
$m(r)$ (dashed line) in cm$^{-2}$ for the $N=5$ CDQR. 
}
\label{fig3}
\end{figure}

We show in Fig.\ \ref{fig2} the sp energy levels as a function of $l$ for
$N=5$ and several $B$ values. Upward (downward) triangles correspond to
spin-up (down) orbitals.
The sp energies are distributed into parabolic-like bands,
each one corresponding to a different value of the radial quantum number
$n$. It can be seen that in most cases, spin-up and -down orbitals
corresponding to the same values of $(n,l)$ are not degenerate due to
the spin magnetization dependence of the exchange-correlation energy 
--we recall that $N$ is odd. Yet, some orbitals still
present this degeneracy, and among them, some are
occupied orbitals, like the $[(0,1)\uparrow,\downarrow]$ ones at $B=1$ T,
or the $[(0,2)\uparrow,\downarrow]$ ones at $B=2$ T.
This can be explained from the different spatial localization of these 
orbitals in the CDQR and the distribution of the spin magnetization
$m(r)=n^{\uparrow}(r)-n^{\downarrow}(r)$, as
shown in Fig. \ref{fig3}: at $B=2$ T, the $l=\pm 2$ sp orbitals are
mostly
localized in the outer ring, in which the local magnetization is fairly
small, whereas at $B=3$ T, the $l=\pm 1$ sp orbitals are mostly
localized in the inner ring, in which the local magnetization is fairly 
small. Eventually, the number of spin-up orbitals is so large
than the $\uparrow,\downarrow$ degeneracy is fully broken 
(this happens for $B>3$ in the displayed cases).

Electron localization in the inner or outer ring induced by 
the magnetic field thus shows up in the KS orbitals, in spite of
having solved the KS equations implicitly assuming a full coherence
regime in which
electrons are allowed to occupy the whole CDQR surface. Yet, they
can be localized in either ring if the
resulting configuration has a lower energy.
It is  relevant
for the analysis of the dipole response to notice that electron
localization is best achieved when, due to the double well structure of
the confining CDQR potential, KS orbitals corresponding to an
intermediate $l$ value are not occupied. In the present case, it happens
for $l=3$. Intuitively, the missing $l$ is the one 
whose KS orbit has a `radius'
similar to that of the maximum of the inter-ring barrier.

A full delocalization situation produces a very regular sp energy
pattern since, due to it, electrons `feel' simultaneously the confining
potential produced by both constituent rings. This happens
e.g. for $B=0$ and other low-$B$ values. The
situation may change at higher magnetic fields. Indeed, Fig. \ref{fig2}
shows that around $B=3$ T, the two lowest parabolic-like bands
tend to cross between $l=2$ and 3. Roughly speaking, each parabolic-like
band arises from one of the constituent rings.
The crossing is quantum mechanically
prevented --level repulsion. The same happens at $B=4$ and 4.5 T. 
A similar effect was found in Ref. \onlinecite{Cli06}, but
described in terms of the inter-ring distance instead of $B$.
Band crossing prevented by level repulsion also appears for the
one-electron CDQR at similar magnetic fields. In this case, spin-up and
-down orbitals are nearly degenerate due to the smallness of the Zeeman
energy.

\begin{figure}[t]
\centerline{\includegraphics[width=6cm,clip]{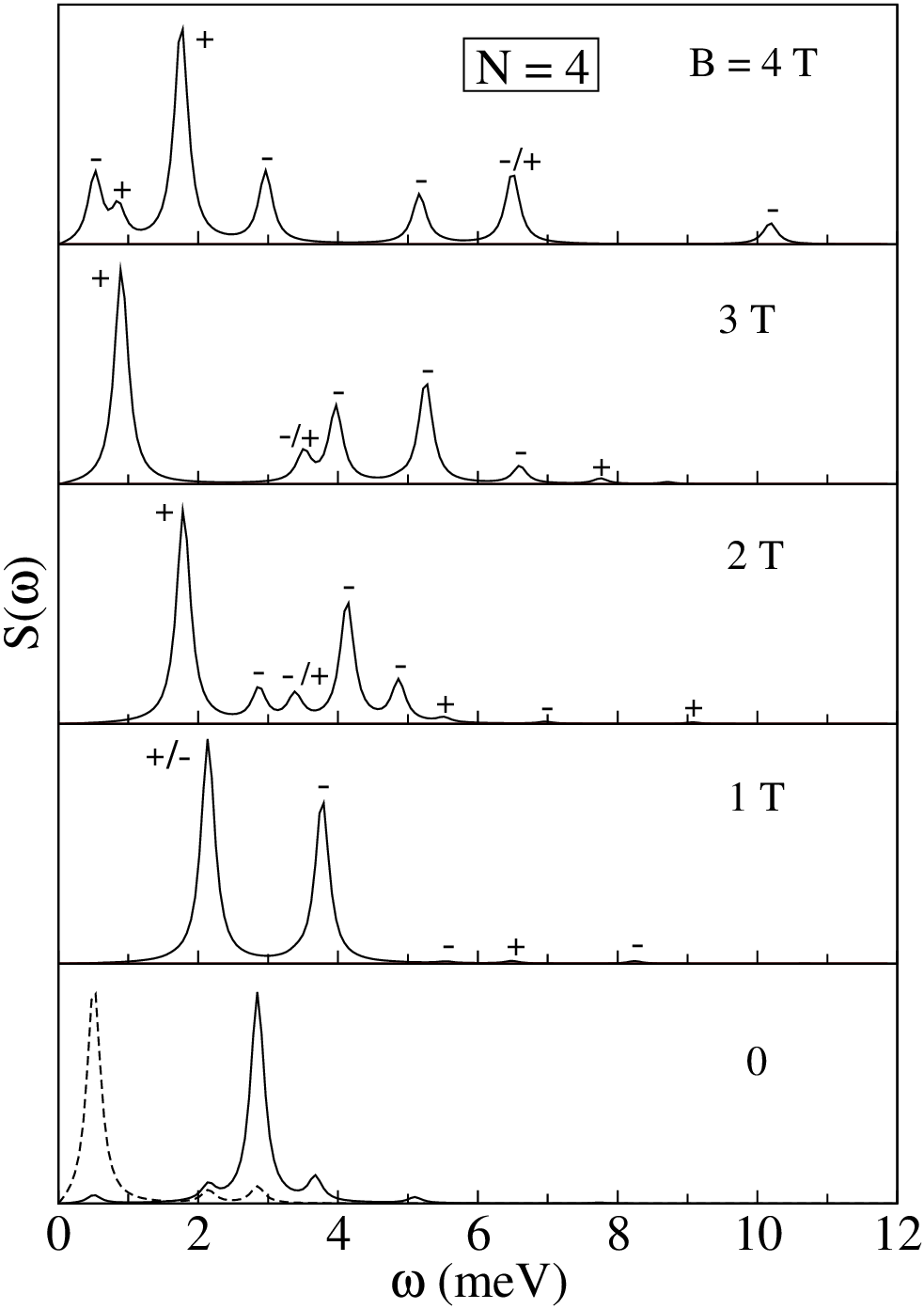} }
\caption{Charge dipole strength (solid lines, arbitrary units) and
spin dipole strength (dashed lines, arbitrary units) for
the $N=4$ CDQR as a function of
the excitation energy (meV) and several $B$ values.
Some structures are superposition of peaks of different polarization,
which is represented by a `$/$' symbol with the polarization of the
more intense peak indicated to the left of the slash.
The intensities are fixed in such a way that for a given $B$, the more
intense peaks in  both channels roughly have the same height.
}
\label{fig4}
\end{figure}

Once the ground state has been determined, the dipole response can be
worked. It is experimentally 
known \cite{Dah93,Lor00} that, for one single quantum ring,  the
dipole spectrum as a function of $B$ consists
of several high- and  low-frequency branches. For large $B$, the
low-frequency branches are identified as edge magnetoplasmons localized at 
the inner and outer ring boundaries. 

In the delocalized regime, the physical pictures for a single quantum
ring and for a CDQR turn out to be similar. The magneto-excitations 
can be classified into bulk (high-energy) and edge (low-energy) modes,
with delocalization producing only two
effective edges: the inner edge of the smaller
ring and the outer edge of the larger ring.
The origin of the high- and low-frequency peaks can be easily understood
from, e.g., the  $B=1$ and 2 T panels in Fig.\  \ref{fig2}.
Bulk, high energy peaks arise from  non spin-flip electronic
excitations mostly
involving $\Delta n=1$, $\Delta l=\pm 1$ changes (inter `Landau level'
excitations), while edge, low energy peaks arise from  non
spin-flip electronic excitations
involving $\Delta n=0$, $\Delta l=\pm 1$ changes (intra `Landau level'
excitations). One thus would expect that the $B$ dispersion of bulk and
edge modes split into two branches each, one corresponding to
$\Delta l=+1$ [circularly polarized $(+)$ excitations], and another to
$\Delta l=-1$ [circularly polarized $(-)$ excitations]. The $(-)$ edge
modes are intra `Landau level'
excitations of the innermost boundary of the double ring system, 
whereas the $(+)$ edge modes
are intra `Landau level' excitations of the outermost boundary. 
The high-energy modes are bulk modes mostly of $(-)$ character, similarly to the
cyclotron mode in quantum dots and wells, and carry much less strength, 
i.e., are less intense. Notice that at $B=0$ the $(\pm)$ excitations are 
degenerate.

\begin{figure}[t]
\centerline{\includegraphics[width=6cm,clip]{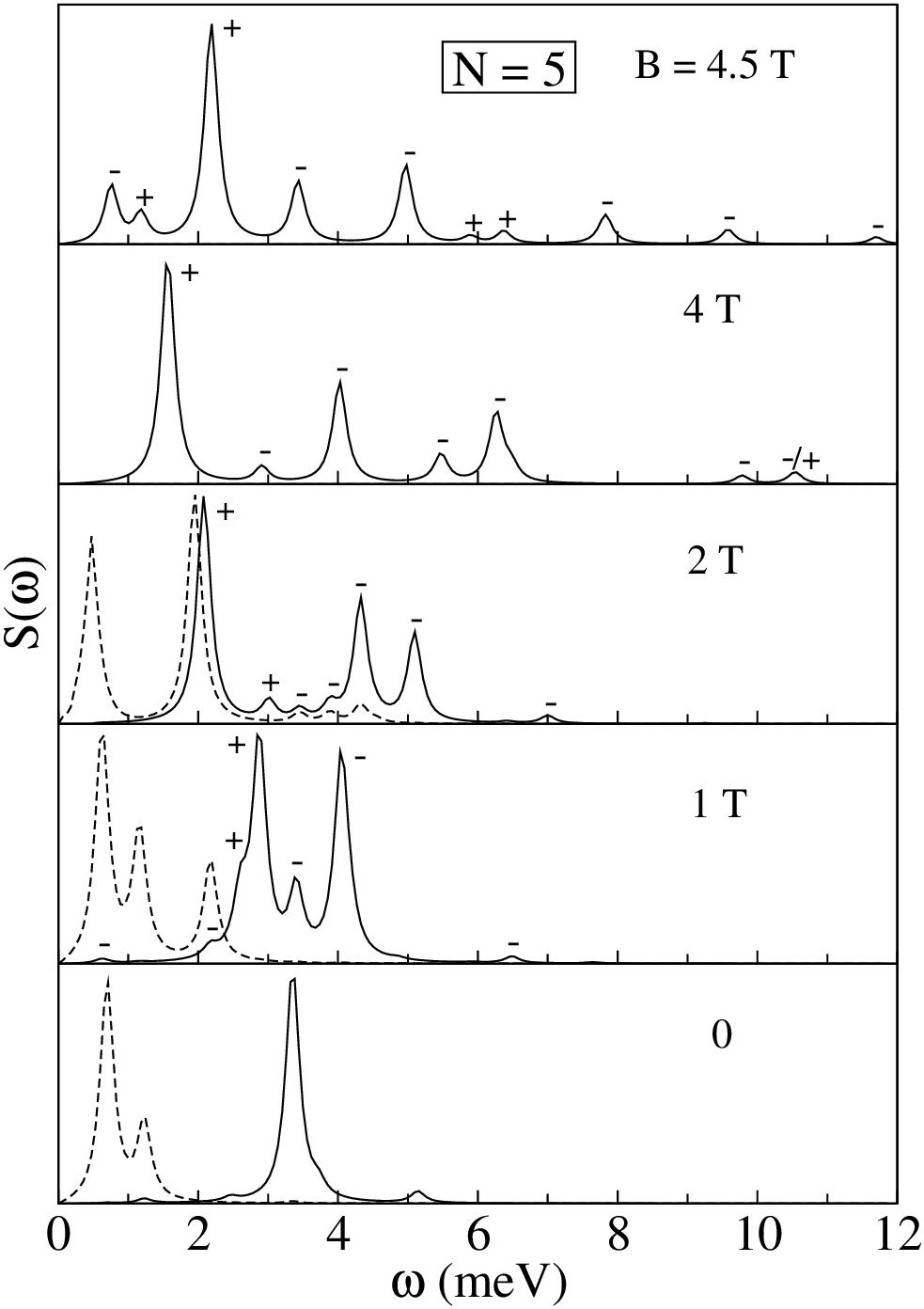} }
\caption{Same as Fig. \ref{fig4} for $N=5$.
}
\label{fig5}
\end{figure}

\begin{figure}[b]
\centerline{\includegraphics[width=6.cm,clip]{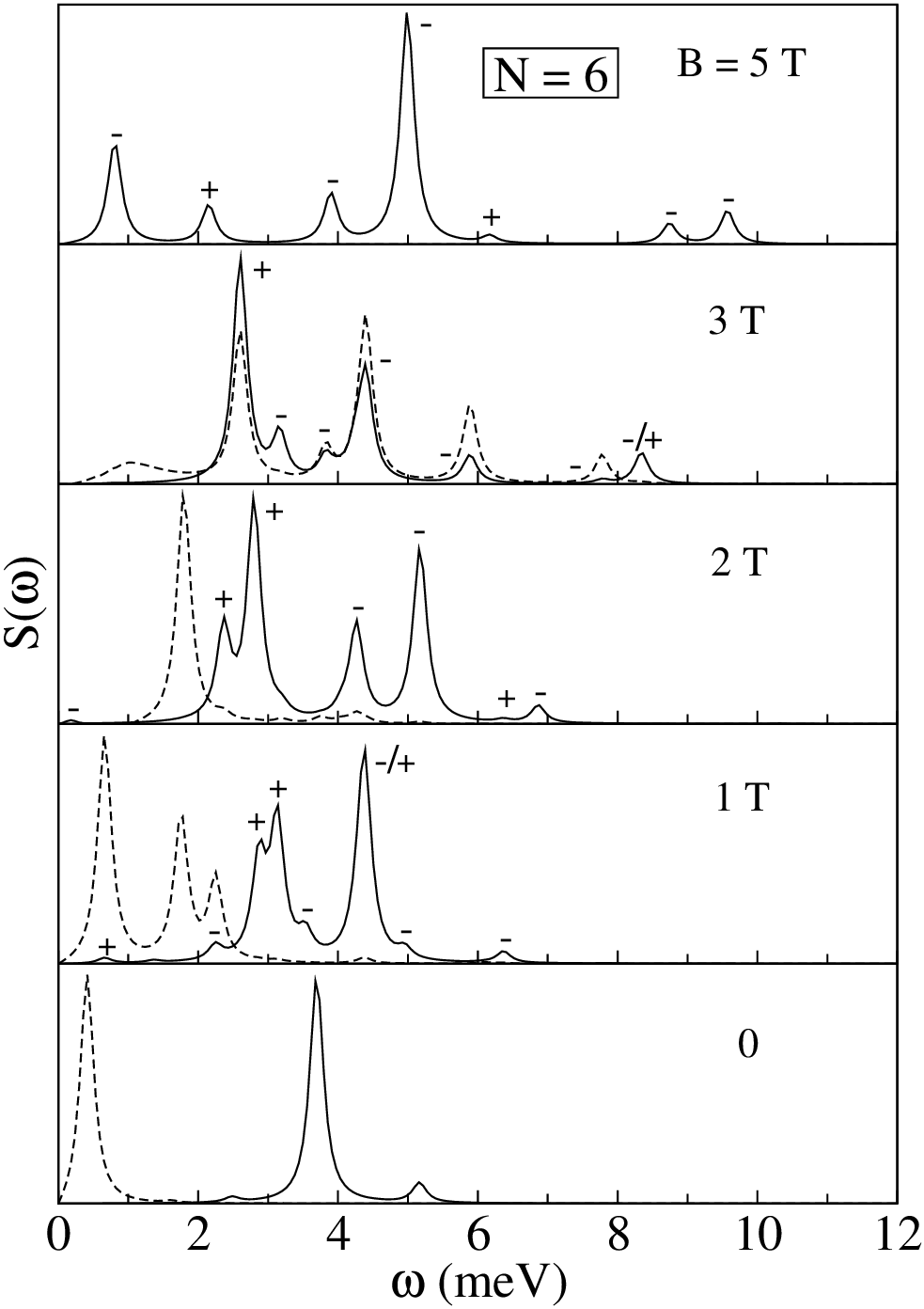} }
\caption{Same as Fig. \ref{fig4} for $N=6$.
}
\label{fig6}
\end{figure}

Examples of this behavior as a function of $B$ are shown in 
Figs. \ref{fig4}-\ref{fig6} for $N=4$, 5, and 6, respectively. The solid
lines in these figures represent the charge dipole strength
$S(\omega)$ in arbitrary
units as a function of the excitation energy. At low $B$ values,
the presence of two edge modes with opposite circular polarizations
can be seen, together with fairly
weak structures arising from $\Delta n=1$ transitions. The figures only 
represent the low energy (up to 12 meV) part of the dipole spectrum.
Some strength is also in the form of low intensity, higher energy modes
arising from $\Delta n=2$ electronic excitations, not shown in these
figures, easily identifiable in plots as those shown in Fig. \ref{fig2}.
Some modes present a fine structure (fragmentation), despite the
tendency of the electron-hole interaction to correlate the 
free-electron excitations, grouping them coherently into few excitation
peaks.

The spin dipole strength is also represented  in these figures in
arbitrary units (dashed lines). Obviously, when the CDQR is fully
polarized both  strengths
coincide and we have plotted only one of them. As a general
rule, charge modes are at higher energies than spin modes because
the electron-hole interaction   is repulsive in the density channel,
as it is essentially determined by the electron-electron Coulomb
interaction, whereas it is  attractive in the spin channel, as it is
determined by the attractive exchange-correlation interaction.
It can be seen from Figs. \ref{fig4}-\ref{fig6} that in some situations,
spin and charge strengths are coupled. This coupling may appear when the
ground state configuration has a non-zero total spin.\cite{Ser99}

Finally, we discuss the case of localized electrons, exemplified here by
the $N=4$, $B=4$ T, the $N=5$, $B=4.5$ T, and the $N=6$, $B=5$ T  CDQR's.
In this regime the response of the system deviates from that of the single 
ring.The fact that the $l=3$ KS orbital is empty
makes it possible to generate additional low energy modes at the inner and outer
boundaries of {\it both} rings. One thus would expect the appearance of
a richer dipole strength, with two series of $(-)$ and $(+)$
polarization
edge modes instead of just one. This is indeed what is shown in  the
corresponding panels of Figs. \ref{fig4}-\ref{fig6}
and it signals the onset of electron localization in the 
strength function.

The present study can be  extended to the case of
many-electron
CDQR and to other multipole excitations, or to incorporate on-plane
wave-vector effects 
for the analysis of Raman
experiments.\cite{Emp01} 
A more realistic description of the CDQR confining potential\cite{Cli06}
demands a three-dimensional approach.\cite{Pi04} 
Work along this line is in progress.

We would like to thank Josep Planelles for useful discussions.
This work has been performed under grants FIS2005-01414 and
FIS2005-02796 from
DGI (Spain) and 2005SGR00343 from Generalitat de Catalunya.
E. L. has been suported by DGU (Spain), grant SAB2004-0091.

\end{document}